\documentclass{pasj00}
\draft

\begin{document}
\SetRunningHead{Author(s) in page-head}{Running Head}
\Received{2006/10/14}
\Accepted{2006/12/07}

\title{Relativistic Effects on the Observed AGN Luminosity Distribution}

\author{Yuan  \textsc{Liu}%
  }
\affil{Physics Department and Center for Astrophysics, Tsinghua
University, Beijing, 100084, China}
\email{yuan-liu@mails.tsinghua.edu.cn}

\author{Shuang Nan  \textsc{Zhang}}
\affil{Physics Department and Center for Astrophysics, Tsinghua
University, Beijing, 100084, China;\\ Key Laboratory of Particle
Astrophysics, Institute of High Energy Physics, Chinese Academy of
Sciences,\\ Beijing, China; \\Physics Department, University of
Alabama in Huntsville, Huntsville, AL35899, USA
}\email{zhangsn@tsinghua.edu.cn} \and
\author{Xiao Ling  {\sc Zhang}}
\affil{Physics Department, University of Alabama in Huntsville,
Huntsville, AL35899, USA}\email{zhangx@email.uah.edu}

%

\KeyWords{galaxies: Seyfert - galaxies: active - galaxies: luminosity function - radiation mechanisms: non-thermal - X-rays: general} 

\maketitle

\begin{abstract}
Recently Zhang (2005) has proposed a model to account for the well
established effect that the fraction of type-II AGNs is
anti-correlated with the observed X-ray luminosity; the model
consists of an X-ray emitting accretion disk coaligned to the
dusty torus within the standard AGN unification model. In this
paper the model is refined by including relativistic effects of
the observed X-ray radiations from the vicinity of the
supermassive black hole in an AGN. The relativistic corrections
improve the combined fitting results of the observed luminosity
distribution and the type-II AGN fraction, though the improvement
is not significant. The type-II AGN fraction prefers non- or
mildly spinning black hole cases and rules out the extremely
spinning case.
\end{abstract}

\section{Introduction}
It is well known that Seyfert Galaxies can be classified as
Seyfert I and Seyfert II. Seyfert I galaxies have both narrow and
broad emission lines in their optical spectra, while Seyfert II
galaxies only have narrow optical emission lines. In the unified
model of Seyfert galaxies (Antonucci 1993), this distinction is
due to the different inclination of our line of sight with respect
to an obscuring torus surrounding the source (Veron-Cetty \& Veron
2000). Therefore, it is expected that the observed luminosity
distribution of these two types should be identical except for the
effects of obscuration by the torus material. However, in several
recent X-ray surveys of AGN (Ueda et al. 2003; Hasinger 2004;
Steffen et al. 2003) it has been found that the fraction of
type-II AGNs is anti-correlated with the observed X-ray
luminosity, such that the luminosity dependent dusty torus model
may be required. However, this observed anti-correlation could be
explained in the framework of unified model if the inclination
angle effects of X-ray radiation are taken into account (Zhang
2005, hereafter Z05). It was pointed by Nayakshin (2006) recently
that the X-ray emission continuum of an accretion disk should be
anisotropic, if the emission is composed of moderately optically
thick magnetic flares (Zhang 2006). Nayaksin (2006) further used
this model to explain the well known observations that the X-ray
continuum and reflected component off the accretion disk are not
well correlated (Markowitz et al. 2003). For simplicity and
focusing on the inclination angle effects, relativistic effects
produced when the X-ray is emitted in accretion disk are ignored
in Z05. However, the relativistic effects certainly exist when the
emission region is close to the black hole. We will consider
relativistic corrections to the observed luminosity distributions
and the type-II AGN fraction in \S2. In this section, we utilize
the combined fitting of the observed luminosity distribution and
the type-II AGN fraction to constrain the spin of supermassive
black holes in the sample. Perhaps both the luminosity dependent
dusty torus model and the above effects operate simultaneously. In
this paper we focus on the effects of inclination angle and
relativistic effects, demonstrating that these effects are also
important. In \S3 we show our conclusion and make some discussion.

\newpage

\section{AGN Luminosity Distribution with Relativistic Corrections}
\begin{figure}
  \begin{center}
    \FigureFile(120mm,120mm){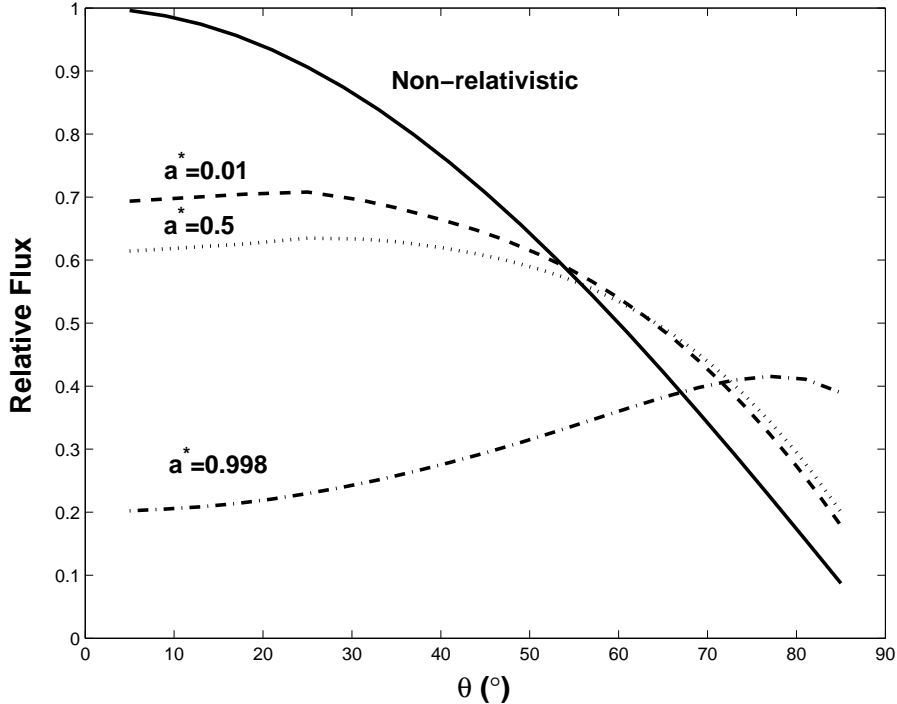}
  \end{center}
  \caption{The observed X-ray flux and inclination angle dependence for non-relativistic
  and relativistic cases in 2-10 keV. The shape strongly depends on the energy range and the spin of black hole
  (Zhang, X. L., et al. 2004). The flux decreases with the value of spin increasing for the face-on cases, since
  the gravitational redshift and transverse Doppler shift (also redshift) become more significant for high
  spinning cases. In contrast the bump for the large inclination angle and high spinning cases is due to the effects of Doppler
  beaming and gravitational focusing (especially for the high energy band concerned here, also see
  L05).
   }\label{fig:1}
\end{figure}

If we consider the inclination angle effects, the type II AGNs
viewed nearly edge-on appear to be less luminous than type I AGNs
viewed nearly face-on for the same intrinsic luminosity. This is
due to the less projected area of the accretion disk, when we
observe it with an inclination angle. Therefore, in the
non-relativistic case when we view an AGN system with an
inclination angle $\theta$, defined as the angle between the
normal direction of the accretion disk (also the spin axis of the
black hole) and the line of sight, the observed luminosity will be
reduced by a factor of $\cos\theta$ compared with the face-on
case, i.e., $\theta=0^\circ$. This projection factor will become
complicated after considering the relativity correction, as shown
in figure \ref{fig:1}. If the limb-darkening effect (Phillips \&
Meszaros 1986) is considered, an additional factor of $(1 + 2\cos
\theta )/3$ to the observed luminosity will be included (Netzer
1987). We assume the orientation of AGNs is isotropic. Therefore,
the probability density of $\theta$ is $\sin\theta $
$(0<\theta<\pi/2)$. With the relation between the flux (i.e. the
luminosity) and $\theta$ (figure \ref{fig:1}), we could obtain the
probability of the observed luminosity , i.e. $f(L)$, for a given
intrinsic luminosity numerically. Then the observed luminosity
distribution is the convolution between $f(L)$ and the intrinsic
luminosity distribution.

The relativistic effects include the Doppler shift and boosting,
the gravitational redshift and light bending. The correction is
calculated by the ray-tracing method (Fanton et al. 1997).  Li et
al. (2005, here after L05) have improved the method by including
the effect of returning radiation and allowing a nonzero torque to
be set at the inner edge of the disk. However, the improvement is
not significant when the free parameter $\eta$ (see L05) is less
than 0.3. Actually, $\eta$ was set to be 0 when the spectra were
fitted by the model in L05. We adopt the standard Keplerian
optically thick and geometrically thin disk (Shakura \& Sunyaev
1973). We assume the disk lies on the equatorial plane of a Kerr
black hole, and the inner edge of the disk is the last stable
orbit and outer edge is 200$r_g$ (beyond this radius the X-ray
radiation is negligible), where $r_g$ is the gravitational radius
of the black hole. For the local spectrum, it contains two
components, a multi-color disk spectrum in all energy bands and a
power law only in 2-10 keV band. The temperature profile of the
blackbody-like component is derived on the Kerr metric (Thorne
1974). However, the temperature of the disk of Seyfert galaxies is
only 50-100 eV or less, so the blackbody-like component is
secondary in the energy range 2-10 keV concerned; we only use its
radiation power to control the distribution of the power law
component. If we assume the total radiation powers of the two
components are equal, then the proportionality of the two
components is about $10^5$ in 2-10 keV band. We assume the value
of the proportionality is 1000 in our calculation, because the
results of luminosity distributions and spectral shapes are not
sensitive to larger values of proportionality of the two
components. The spectral index is assumed to have a form $\alpha =
- 0.45 + (r/r_g )^{1/4} /1.2$ , in order to mimic the shape of the
observed power law after the relativistic corrections. We stress
that this spectral model is only phenomenological, because
currently no self-consistent physical model of accretion disk
emission (because the nature of the inferred X-ray emitting
``corona" is currently unknown and remains a controversial issue)
is available to account for the observed AGN X-ray spectra in
details, without invoking some kind of assumptions and/or
parameterizations.  However, our main results about the total flux
are not sensitive to the exact form of the phenomenological model
used here, because our main goal is to re-produce the observed
X-ray spectra, which is of course also the goal of any physically
self-consistent accretion disk emission model.

The intrinsic luminosity distribution, referred to as the AGN
luminosity before correcting for the inclination angle effect and
relativistic effects, is assumed of a broken power-law shape,
i.e., $N \propto L_X^\alpha  $, where $L_X$ is in units of erg/s,
$\alpha=0.25$ for $10^{42.75 + x}  < L_X  \le 10^{44.9 + x} $
 and  $\alpha=-0.7$ for $10^{44.9 + x}  < L_X  < 10^{47 + x} $
; these parameters are determined by matching the data with the
model predictions (Z05). The parameter $x$ is changed for
different cases to obtain the best agreement (i.e. the minimum
$\chi^2$ values) with the data. We use luminosity distribution
instead of luminosity function which is usually used in studying
AGN. Indeed given the instrument sensitivity and sky exposure, the
luminosity distribution can be derived uniquely from any assumed
luminosity function. Our goal is to understand the effects of
inclination angle and relativistic effects to the observed
luminosity of AGNs, rather than attempting to understand the
physical origin of the intrinsic luminosity function. Therefore,
we may study either luminosity function or luminosity
distribution. Since complex modeling is required to obtain the AGN
luminosity function, but the luminosity distribution is obtained
directly from data, we choose to focus our study on luminosity
distribution, in order to understand the effects of inclination
angle and relativistic effects to the observed luminosity of AGNs.

\begin{figure}
  \begin{center}
    \FigureFile(120mm,120mm){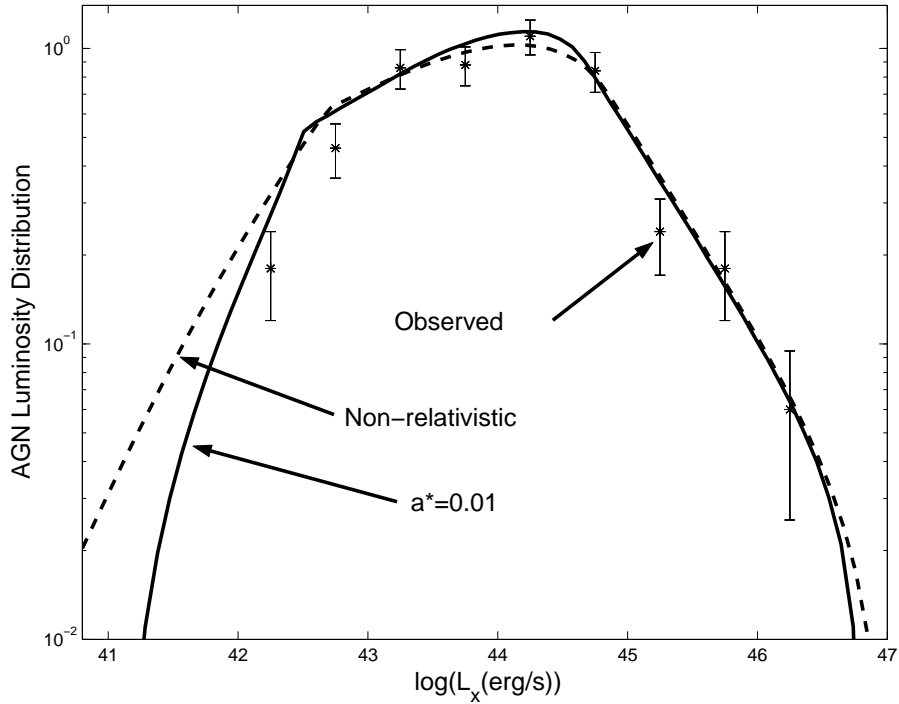}
  \end{center}
  \caption{The predicted AGN luminosity distributions for non-relativistic and $a^*=0.01$ cases (2-10 keV). The
  observed luminosity distribution (after absorption corrections) of AGNs (Ueda et al. 2003) agrees with the
  predicted apparent (observed) luminosity defined as  $L_X  = F_X 4\pi D_L^2 $
, where  $F_X$ is the observed X-ray flux
  after absorption corrections and  $D_L$ is the luminosity distance of the AGN.}\label{fig:2}
  \end{figure}

\begin{figure}
  \begin{center}
    \FigureFile(120mm,120mm){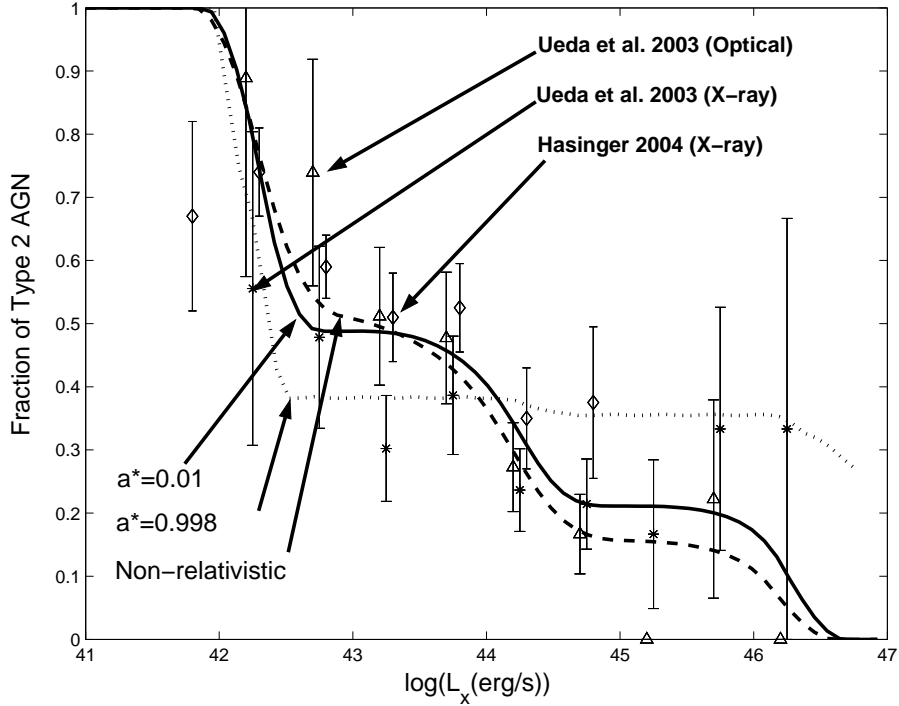}
  \end{center}
  \caption{Type-II AGN fraction as function of the observed apparent X-ray luminosity (after absorption
  corrections) for non-relativistic, $a^*=0.01$ and 0.998 cases. The data points shown by diamonds and triangles
  are shifted horizontally by 0.05 and -0.05 respectively for displaying clarity. Because the three different
  groups of type-II AGNs, i.e., optical and X-ray type-II AGNs from Ueda et al. (2003) and X-ray type-II AGNs
  from Hasinger (2004) may have slightly different definitions in terms of the dividing inclination angle between
  type-I and type-II AGNs. As shown in this figure, the $a^*=0.998$ case cannot produce the decreasing trend of
  type-II AGN fraction in the high luminosity band ( $L_X  > 10^{43.5} \;{\rm{erg/s}}$
). }\label{fig:3}
  \end{figure}

\begin{figure}
  \begin{center}
    \FigureFile(120mm,120mm){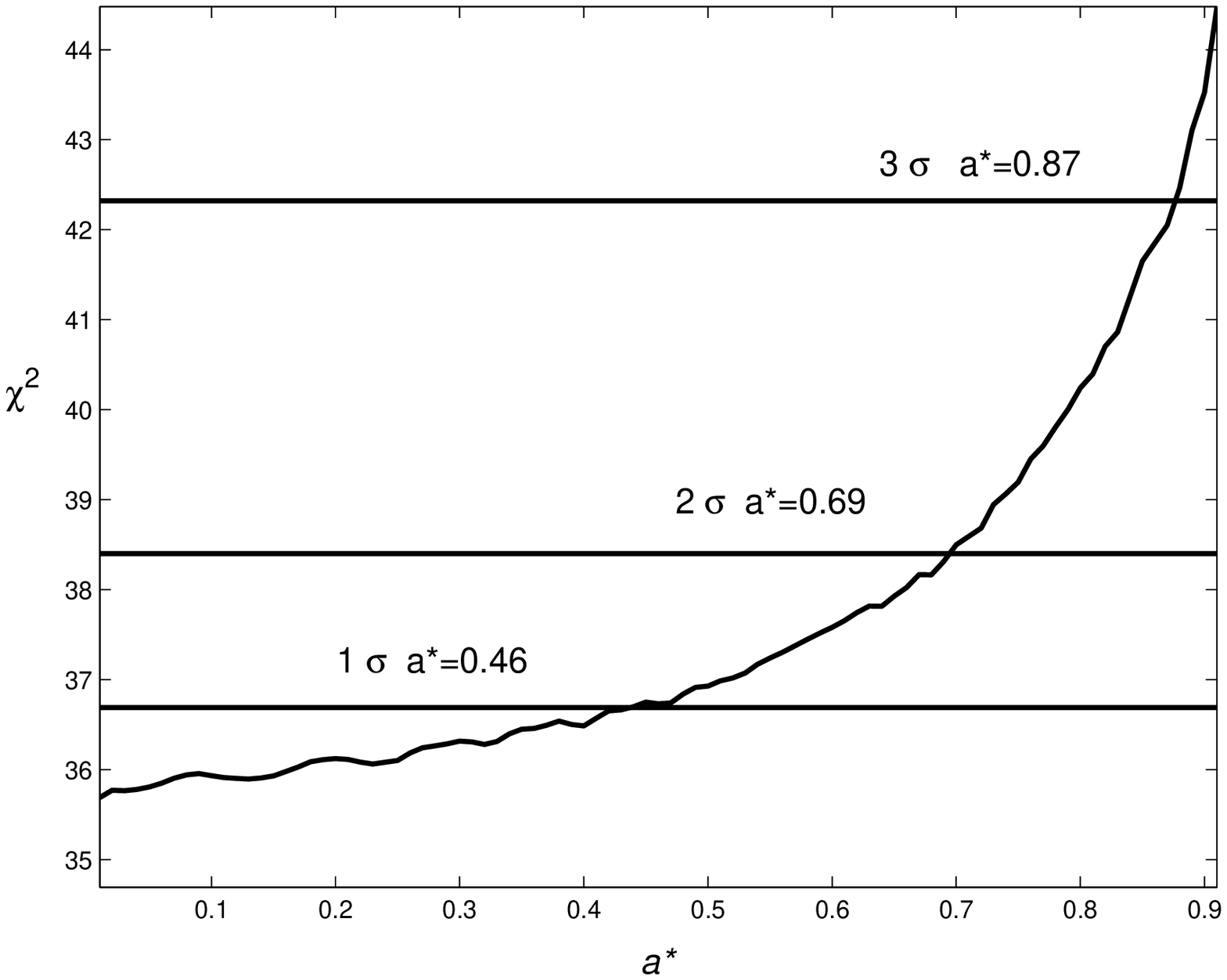}
  \end{center}
  \caption{ The  $\chi _{{\rm{total}}}^2 $
 curve by fitting the observed luminosity distribution and the type-II AGN fraction as
  function of apparent X-ray luminosity. Both $x$ and $\theta_c$ are also adjusted to obtain the minimum  $\chi _{{\rm{total}}}^2 $
 for
  each value of $a^*$. }\label{fig:4}
  \end{figure}

We fit the observed luminosity distribution and the type-II AGN
fraction at the same time in order to distinguish between
different models. The total  $\chi _{{\rm{total}}}^2 $
 is defined as  $\chi _{\rm{L}}^2  + \chi _{\rm{F}}^2 $
, in which   $\chi _{{\rm{L}}}^2 $ and $\chi _{{\rm{F}}}^2 $ are
obtained with the observed luminosity distribution and the type-II
AGN fraction respectively. The minimum value of $\chi
_{{\rm{total}}}^2 $
 for non-relativistic case is 38.17 ($x =  - 0.01_{ - 0.05}^{ + 0.06} $
, the dividing inclination angle between type II and type I AGNs,
$\theta _c  = {65.5^ \circ} _{ - 1.0^ \circ  }^{ + 1.5^ \circ  } $
, the degree of freedom is 30, i.e. DOF=30); for the relativistic
case, the minimum value of $\chi _{{\rm{total}}}^2 $
 is obtained when $a^*=0.01$ ($\chi ^2  = 35.69$
 , $x =  - 0.03_{ - 0.03}^{ + 0.04} $
 , $\theta _c  = {66.5^ \circ}  _{ - 1.0^ \circ  }^{ + 1.5^ \circ  } $
 , DOF=29). Although the results of the relativistic case are better
 than the non-relativistic case, the improvement is not significant. The significance of improvement given by $F$-test
 is 83.4\%. We show the results of the observed luminosity distribution and the type-II AGN fraction in figure \ref{fig:2} and \ref{fig:3}
 respectively. With the fitting result, we could determine the upper
limits of $a^*$ are 0.46, 0.69 and 0.87 corresponding
 to 1, 2 and 3$\sigma$, respectively (figure \ref{fig:4}). For the extremely spinning case ($a^*$=0.998), the minimum
 value of  $\chi _{{\rm{F}}}^2 $
 is 54.02 ( $x=0.27$, $\theta _c  = 68.0^ \circ  $
), corresponding to a probability less than $10^{-10}$ . On the
other hand, the upper limit of $a^*$
 determined by $\chi ^2 $
statistic is 0.75 with 90\% confidence, which is consistent with
the results above.

In summary, the combined fitting result of the observed luminosity
distribution and the type-II AGN fraction favors non- or mildly
spinning black hole cases; the extremely spinning case is ruled
out with high confidence. The dividing inclination angle between
type II and type I AGNs is between 60 and 70 degrees, slightly
different from that of 68 and 76 degrees in Z05 and in slightly
better agreement with the range of inclination angles of type I
AGNs determined by Wu \& Han (2001).

\section{Conclusion and Discussion}

The main results of this paper are: (1) The relativistic
corrections improve the combined fitting results of the observed
luminosity distribution and the type-II AGN fraction, though the
improvement is not very significant. (2) The fitting results
prefer non- or mildly spinning black hole cases, and the extremely
spinning case is ruled out with high confidence.

The result of type-II AGN fraction shows that the majority of
black holes in the sample cannot be extremely spinning, especially
for the high luminosity ones (as shown in figure \ref{fig:3}).
Several authors have investigated the spin of supermassive black
holes in AGNs  in both theoretical and observational aspects. A
considerable fraction of high luminosity AGNs in the sample in
Ueda et al. (2003) and Hasinger (2004) are high redshift QSOs ($z
= 1\sim3 $). For the QSOs, some recent calculations about the
evolution of the spin of supermassive black holes (Shapiro 2005;
Volonteri et al. 2005) find that most black holes are rapidly
spinning by $z\sim5$. In their calculations, the accretion rates
have been set equal to the Eddington accretion rate all the time.
We have also investigated how the evolution of accretion rate
impacts the evolution of the spin of supermassive black holes. We
have found that the spin of black hole should have approached the
maximum value before the accretion rate has declined significantly
after the QSO formation, which is most likely a merger event (Yu
et al. 2005). It is therefore clear that black holes should become
extremely spinning in the QSO stage very rapidly, provided that
there are not other mechanisms that could reduce the spin.
However, extraction of black hole's spin energy is believed to be
the main mechanism in powering relativistic jets from supermassive
black holes and thus preventing it from becoming rapidly spinning
in the QSO stage (Koide et al. 2002). In fact it is highly
possible that QSOs may lose their spin energy very rapidly, so
that their black holes only remain in the highly spinning state
shortly, i.e., the high accretion rate stage, agreeing with the
fact that most QSOs are radio quiet. In addition, the supermassive
black holes could grow up while keeping the spin low by chaotic
accretion (King \& Pringle 2006). As a result of the above
possible mechanisms, the average spin of QSOs is likely to be only
moderate. For Seyfert galaxies, if they have not undergone a
merger event recently (Grogin et al. 2005) and then maintained low
accretion rate, their spin could also be low or mild.

The spin of a black hole can also be inferred from the radiation
efficiency. The radiation efficiencies corresponding to the $1$,
$2$ and $3\sigma$ upper limit of the spin deduced from our results
are about 8\%, 10\% and 15\%, respectively. Yu \& Tremaine (2002)
found $\eta\geq0.1$ of optical selected quasar, while the results
from the cosmic X-ray background indicate $\eta\geq0.15$ (Elvis et
al. 2002). Higher values of $\eta$ (about 30\%-35\%) have been
obtained by Wang et al. (2006) from a large sample of quasars
selected from the Sloan Digital Sky Survey. However, it may be
reduced by a factor $\sim2$ due to the uncertainty of the black
hole mass (Wang et al. 2006). Therefore, our results are not in
conflict with these previous results, but instead provide useful
constraints.
\\

SNZ acknowledges partial funding support by the Ministry of
Education of China, Directional Research Project of the Chinese
Academy of Sciences and by the National Natural Science Foundation
of China under project no. 10521001.

\newpage



\begin{thebibliography}{}
\bibitem[Author(2001)]{key-1}
Antonucci R. 1993, ARA\&A, 31, 473
\bibitem[Author(2001)]{key-1}
Elvis, M., Risaliti, G., \& Zamorani, G. 2002, ApJ, 565, L75
\bibitem[Author(2001)]{key-1}
Fanton, C. et al. 1997, PASJ, 49, 159
\bibitem[Author(2001)]{key-1}
Grogin, N. A. et al. 2005, ApJ, 627, L97
\bibitem[Author(2001)]{key-1}
Hasinger, G. 2004, Nucl. Phys. B Proc. Suppl., 132, 86
\bibitem[Author(2001)]{key-1}
King, A. R., \& Pringle, J. E. 2006, MNRAS, in press
 (astro-ph/0609598)
\bibitem[Author(2001)]{key-1}
Koide, S. et al. 2002, Science, 295, 1688
\bibitem[Author(2001)]{key-1}
Li, L. X. et al., 2005, ApJS, 157, 335
\bibitem[Author(2001)]{key-1}
Markowitz, A., Edelson, R., \& Vaughan S., 2003, ApJ, 598, 935
\bibitem[Author(2001)]{key-1}
Nayakshin, S. 2006, submitted to MNRAS Letters (astro-ph/0611347)
\bibitem[Author(2001)]{key-1}
Netzer, H. 1987, MNRAS, 225, 55
\bibitem[Author(2001)]{key-1}
Phillips, K. C., \& Meszaros, P. 1986, ApJ, 310, 284
\bibitem[Author(2001)]{key-1}
Shakura, N. I., \& Sunyaev, R. A. 1973, A\&A, 24, 337
\bibitem[Author(2001)]{key-1}
Shapiro, S. L. 2005, ApJ, 620, 59
\bibitem[Author(2001)]{key-1}
Steffen, A. T., Barger, A. J., Cowie, L. L., Mushotzky, R. F., \&
Yang, Y. 2003, ApJ, 596, L23
\bibitem[Author(2001)]{key-1}
Thorne, K. S. 1974, ApJ, 191, 507
\bibitem[Author(2001)]{key-1}
Ueda, Y., Akiyama, M., Ohta, K., \& Miyaji, T. 2003, ApJ, 598, 886
\bibitem[Author(2001)]{key-1}
van Putten, M.H.P.M., \& Levinson, A. 2002, Science, 295, 1874
\bibitem[Author(2001)]{key-1}
Veron-Cetty, M. P., \& Veron, P. 2000, A\&ARv, 10, 81
\bibitem[Author(2001)]{key-1}
Volonteri, M. et al. 2005, ApJ, 620, 69
\bibitem[Author(2001)]{key-1}
Wang, J. M. et al. 2006, ApJ, 642, L111
\bibitem[Author(2001)]{key-1}
Wu, X., \& Han, J. L. 2001, ApJ, 561, L59
\bibitem[Author(2001)]{key-1}
Yu, Q., \& Tremaine, S. 2002, MNRAS, 335, 965
\bibitem[Author(2001)]{key-1}
Yu, Q., Lu, Y., \& Kauffmann, G. 2005, ApJ, 634, 901
\bibitem[Author(2001)]{key-1}
Zhang, S. N. 2005, ApJ, 618, L79 (Z05)
\bibitem[Author(2001)]{key-1}
Zhang, S. N. 2006, Invited Discourse, the 26th IAU General
Assembly, Prague, Czech Republic
\bibitem[Author(2001)]{key-1}
Zhang, X. L., Zhang, S. N., Feng, Y. X., \& Yao, Y. S. 2004, HEAD,
8.4001Z


\end{thebibliography}
\end{document}